# Statistics students' identification of inferential model elements within contexts of their own invention

Matthew D. Beckman

*Pennsylvania State University, 326 Thomas Building, University Park, PA 16802*
mdb268@psu.edu

Robert delMas

*University of Minnesota, 56 East River Road, Minneapolis, MN 55455*

**Abstract**

Statistical thinking partially depends upon an iterative process by which essential features of a problem setting are identified and mapped onto an abstract model or archetype, and then translated back into the context of the original problem setting (Wild and Pfannkuch 1999). Assessment in introductory statistics often relies on tasks that present students with data in context and expects them to choose and describe an appropriate model. This study explores post-secondary student responses to an alternative task that prompts students to clearly identify a sample, population, statistic, and parameter using a context of their own invention. The data include free-text narrative responses of a random sample of 500 students from a sample of more than 1600 introductory statistics students. Results suggest that students' responses often portrayed sample and population accurately. Portrayals of statistic and parameter were less reliable and were associated with descriptions of a wide variety of other concepts. Responses frequently attributed a variable of some kind to the statistic, or a study design detail to the parameter. Implications for instruction and research are discussed, including a call for emphasis on a modeling paradigm in introductory statistics.

# 1. Introduction

"Models are one of the most important and yet least understood ideas in an introductory statistics course" (Garfield and Ben-Zvi 2008). The use of statistical modeling to connect data, chance, and context engages students in each facet of statistical thinking, reasoning, and literacy. Statistical thinking has been described in part to concern comprehension of *how, when,* and *why* a statistical framework informs an inquiry (Ben-Zvi and Garfield 2005). In learning and cognition research, cognitive transfer—or simply 'transfer'—is an important mechanism by which students accomplish this type of comprehension. Singley and Anderson (1989, p. 1) defined transfer to concern "how knowledge acquired in one situation applies (or fails to apply) in other situations."

During a prior study, Beckman (2015) evaluated student responses to an assessment tool designed to measure emergent statistical thinking and associated transfer outcomes administered near the end of their introductory statistics course. While not a primary focus of the study, Beckman remarked on his impression that students attributed a variety of concepts to the parameter and recommended future research on the subject. This paper describes a new study that examines students' recognition of inferential model structures using extant data from the Beckman (2015) study. The data are free-text narrative responses produced by introductory statistics students largely from postsecondary institutions across the United States to a task designed to reveal students' ability to differentiate concepts of sample, population, statistic, and parameter in a context of their own invention.

# 2. Literature Review

The term "model" takes on different meanings in mathematics, statistics, mathematics education, and statistics education (e.g., Garfield and Ben-Zvi 2008;

Graham 2006; McCullagh 2002). For the purpose of this study, “statistical model” or just “model” can be understood to mean a statistical approximation intended to reflect or describe the underlying structure of a data generating process. This aligns closely to an interpretation proposed by Graham (2006) as an application of mathematical modeling for use in statistics, while still preserving flexibility to encompass a variety of statistical models in common use. For example, more advanced models including mixed effects or multivariate varieties as well as comparatively simpler univariate models could all be reasonably characterized as statistical approximations designed to extract or explain the underlying structure of a data generating process. There is some inconsistency in both literature and common usage as to whether the “model” is ascribed to the theoretical ideal or the empirical realization (Graham 2006).

Pfannkuch et al. (2018; Wild and Pfannkuch 1999) emphasized the dynamic interplay between available data, context, and model in an investigative cycle that iteratively revisits and refines understanding of each aspect. Pfannkuch et al. (2018) identified three levels of model recognition that emerge from this cycle. One level requires recognition of an archetypical model pertinent to the situation at hand; a second involves comparison of real data to a model derived mathematically; a third pertains to recognition that an underlying model likely exists although it is not known (Pfannkuch et al. 2018). Since it is difficult for a novice to recognize problem structure and sufficiently generalize cognitive elements on their own, the educator must engage strategic methods to facilitate the desired abstraction (Reed et al. 1985). A tension emerges as teaching and learning emphasize abstraction of subject matter while scenarios in the real-world require contextualized reasoning (Bransford et al. 2000). Information understood in the abstract must be usable for a particular situation (Singley and Anderson 1989) in order for the necessary shuttling between context domain and the archetypical model to occur.

Assessment of statistical inferential thinking typically involves tasks that present students with data in context and ask them to identify an appropriate statistical model. An alternative approach is to present students with the abstract concepts and then prompt them to respond with an appropriate context. For example, Chance (2002) describes the following assessment item from Rossman and Chance (2001):

> The underlying principle of all statistical inference is that one uses sample statistics to learn something… about the population parameters. [Write] a short paragraph describing a situation in which you might use a sample statistic to infer something about a population parameter. Clearly identify the sample, population, statistic, and parameter in your example. Be as specific as possible, and do not use any example which we have discussed in class.

The above item (hereafter the RC task) is unique in the overt way that it reverses the direction of conventional assessment tasks. As a result, the item is well-suited to assess recognition of an archetypical model pertinent to the situation at hand as described by Pfannkuch et al. (2018). With respect to statistical thinking, this item allows the assessment of students' mental habits with respect to their problem-solving approach and whether they focus on critical aspects of the problem (Chance 2002).

There is a dearth of research that directly assesses college students' ability to distinguish sample from population or statistic from parameter or on students' ability to recognize models *a priori*. Studies of primary and secondary students indicate that understanding the distinction between samples and a population improves with grade level (Lavigne and Lajoie 2007; Watson and Kelly 2005; Watson and Moritz 2000) and instruction (Meletiou-Mavrotheris and Paparistodemou 2015). Even teachers sometimes conflate sample and population (Makar and Rubin 2009; Pfannkuch 2006). There is also evidence that college students and research practitioners misunderstand concepts related to statistical significance tests (delMas et al. 2007; Haller and Kraus 2002; Vallecillos 1999; Well et al. 1990; Williams 1999).

Kaplan et al. (2009, 2010) studied the related issue of lexical ambiguity such that students ascribe inappropriate meaning to terms with a precise technical definition in statistics, especially when the statistical meaning of a word is different from common usage or understanding. Kaplan et al. (2009) found a large variety of meanings for the words “average” and “spread” among students in their studies, whereas there was more convergence in prior meanings for the words “confidence,” “random” and “association.” They argued that the latter three words might be easier to address through instruction than the former two words (Kaplan et al. 2009, 2010). Kaplan and Rogness (2018) specifically explored lexical ambiguity related to the word pair ‘parameter’ and ‘statistic.’ They found that ‘parameter’ was often interpreted to mean a “rule or guideline to be followed” (p. 8) and reported that 60% of students interpreted ‘statistic’ to be a calculated value or variable that was not clearly linked to the sample or population.

### 2.1 Research Questions

We lack knowledge about how students who complete an introductory statistics course perform in model recognition tasks when asked to distinguish foundational concepts related to inferential modeling. To address this, the current study was guided by two research questions: (1) When asked to provide a context for statistical inference of their own devising, what aspects of the context do students attribute to concepts of sample, population, statistic and parameter? (2) Is the impression cited in previous research that students incorrectly attribute a variety of concepts to the parameter warranted in an independent sample and, if so, what are these concepts?

## 3. Methods

### 3.1 Assessment Instrument

An assessment called the Introductory Statistics Understanding and Discernment Outcomes (I-STUDIO) instrument was developed and revised through an iterative process including expert feedback and piloting as part of a PhD dissertation that studied cognitive transfer outcomes for undergraduate introductory statistics students as part of a prior study (see Beckman 2015). A modification of the RC task was included among 7 open-ended assessment tasks that comprise the I-STUDIO assessment tool.

### 3.2 The RC task

The RC task was modified for I-STUDIO as follows:

> An underlying principle of all statistical inference is that one uses sample statistics to learn something about the unknown population parameters. Demonstrate that you understand this statement by describing a realistic scenario in which you might use a sample statistic to infer something about a population parameter. For the context of your example, clearly identify:
>
> - the research question for your scenario,
> - the sample,
> - the population,
> - the statistic, and
> - the parameter.
>
> Be as specific as possible, and do not use any example that was discussed in your statistics course.

The modification more clearly emphasizes the desired components of a complete response. The sample, population, statistic, and parameter are the target content elements for the purposes of this study. The added requirement to state a research question provided a context to facilitate assessment of the target content elements.

### 3.3 Sample

The I-STUDIO instrument was field tested by 14 instructors in 29 class sections for 16 unique courses at 15 different institutions (one instructor had dual affiliation). All instructors were asked to offer course credit (e.g. homework, extra credit) to incentivize effort. Free-text responses to I-STUDIO were collected using a web-based assessment platform. A total of 1,975 students submitted responses, with 1,935 students giving consent and 1,622 students completing the RC task. The maximum enrollment aggregated across all participating institutions based on published enrollment figures was estimated at 2,265. The usable response rate was estimated between 71.6% and 82.1% depending on whether the maximum enrollment estimate or total responses are used as the denominator. A random sample of 500 responses to the RC task were evaluated for the current study.

## 3.4 Scoring

Initially, a random sample of 10 responses was selected and scored in order to initialize a scoring rubric. The order of all 1,622 responses were randomized to reduce the impact of carryover effects for students who may have been in the same class, institution, etc. The first 500 of the 1,622 randomized student responses to the RC task were evaluated. The 10 responses used to initialize the scoring rubric were not overtly included, but no attempt was made to exclude them from the sample. Scoring criteria determined whether or not each response represented a genuine attempt and then assessed content knowledge by identifying the concept to which the student attributed each of the target content elements (i.e. sample, population, statistic, parameter).

### *3.4.1 Scoring tools*

Scoring and data analysis were performed using R version 3.3.3 (R Core Team 2017). The first author developed a scoring guidance document and user interface in order to streamline the scoring effort and reduce errors. The user interface showed one student response in the console at a time, and then displayed a sequence of prompts to the scorer (Figure 1).

The first author scored all responses. A scoring guidance document was visible at all times when scoring (see the online supplement). The scoring guidance document showed each prompt displayed by the user interface and provided interpretations of common scoring choices. Importantly, the document contained a table of additional scoring codes in use by the scorer to support consistency in coding.

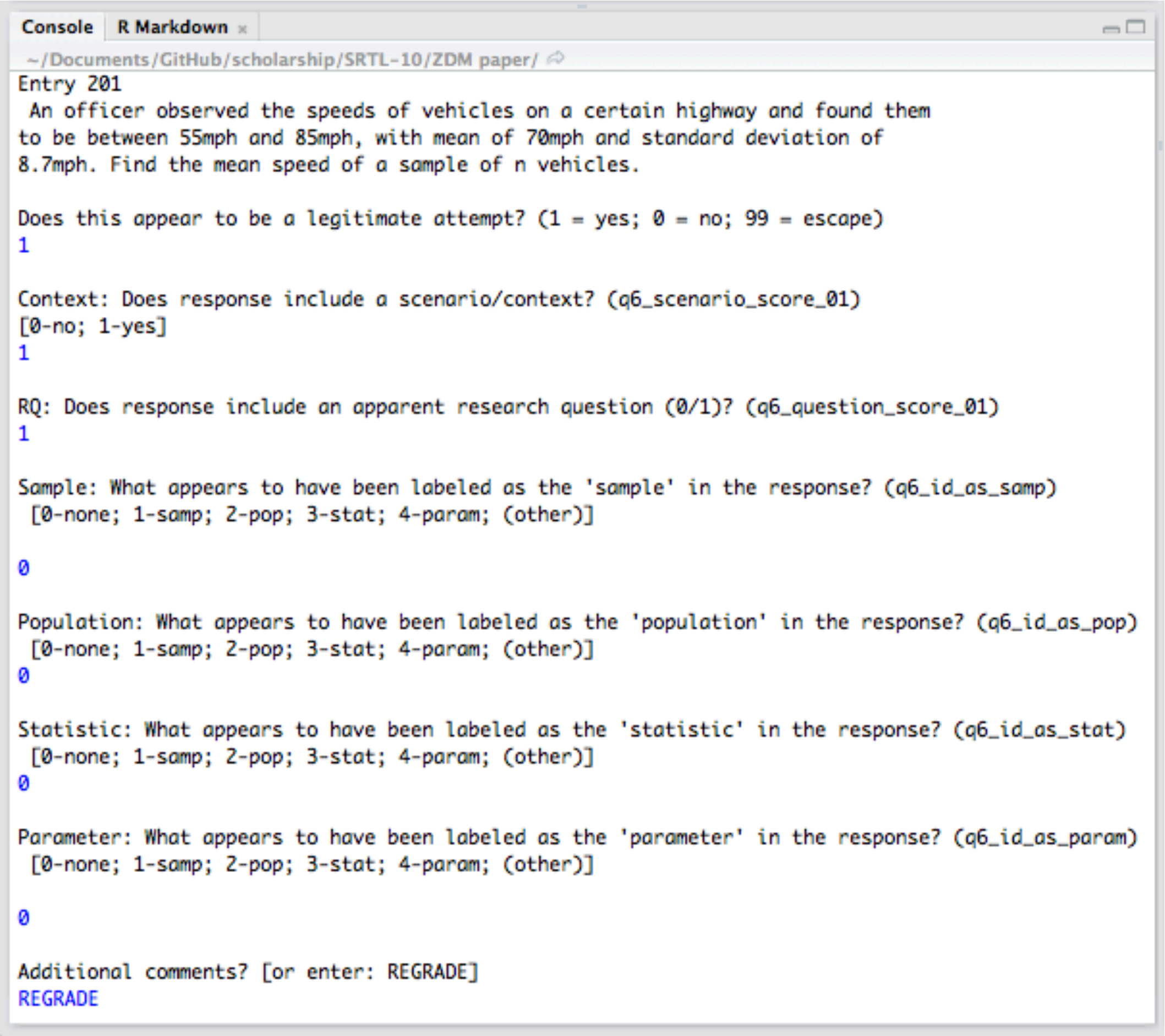

Figure 1. Screenshot of the user interface for scoring.

### *3.4.2 Evaluating student interpretations of inferential modeling elements*

The primary outcome of interest assessed by the RC task concerns concepts to which the student attributes each of four elements essential to statistical inference (i.e. sample, population, statistic, parameter) within a context of their own devising. Scoring codes were stored as free text, so results were parsed using regular expression matching and alternative scoring codes were added as unique cases emerged. For example, the scoring code of "partial" was added to indicate partial credit for cases where it was difficult to determine whether the student demonstrated a complete understanding of the concept.

Periodically, all scoring codes that did not conform to codes defined in the scoring guidance document were scrutinized and consolidated as appropriate. All affected responses were revisited to verify that the updated code accurately represented each response before any code was modified.

The analysis considered those attempts that were found to provide a context or scenario of some kind as part of the solution. Such responses indicated strong evidence that the student attempted to provide a legitimate response that he or she intended to be scored.

### *3.4.3 Coding and scoring responses*

Table 1 shows a selection of verbatim responses to the RC task intended to illustrate several typical examples. The ID number simply represents the random order used for scoring purposes. The content element requested, code attributed to student response, and comments are included to highlight the scoring method.

Response ID 38 earned full credit. Response ID 148 identified appropriate examples for the sample and population but attributed the rephrased research question to the statistic and a study design detail to the parameter. Response ID 177 attributed a variable to the statistic, and a study design detail to the parameter. Response ID 175 earned partial credit for the population since the response demonstrated a pertinent understanding of the concept but was judged not sufficiently clear or precise to warrant full credit. Response ID 175 also reversed the labels of 'statistic' and 'parameter.' Response ID 231 was coded as partial credit for the population. One could argue that the response demonstrated a reasonably clear understanding of population, but it did not explicitly link the description to the "population." Response ID 231 did not identify the statistic or parameter.

Table 1. Coding examples of verbatim student responses to the RC task

| ID | Verbatim response to RC task | Content element requested by RC Task | Code attributed to student response | Coding comments |
|---|---|---|---|---|
| 38 | one wants to find out how much all [country] university students spend on lunch.<br>it is hard to calculate this of the all the [country] students, hence, the whole population. Therefore, one could take a sample, for example, by gathering this data from 80 students from every [country] university. One could take the mean of this data, which would be a statistic, which is used to estimate the true amount of money which [country] students spent on lunch, this later value is called the parameter. | Sample | Sample | This response earned full credit for clearly identifying the sample, population, statistic, and parameter. |
| | | Population | Population | |
| | | Statistic | Statistic | |
| | | Parameter | Parameter | |
| 148 | Research question: does getting paid for school have an effect on grades?<br>sample: 30 students out of the each class of [institution's] undergraduate population<br>population: [institution] undergraduate population<br>Statistic: unsure what this means...maybe if grades were higher for those paid than unpaid<br>parameter: one semester | Sample | Sample | This response identifies an appropriate sample and population. The statistic states a result that would affirm the research question, and the parameter is attributed to a study design detail (duration) |
| | | Population | Population | |
| | | Statistic | RQ (i.e., restatement of research question) | |
| | | Parameter | Study design detail | |
| 177 | research question: average length of stay of person recovering from gastric bypass surgery, the sample would be from three different hospitals located across the [country], the population would be males and females who are recovering from gastric bypass surgery the statistic is the length of days and the parameter is 30-45 years old, overweight. | Sample | Sample | This response characterizes the whole response as a "research question." The response reasonably describes the sample and population, yet the statistic is attributed to a variable (length of days) and the parameter appears to describe participant selection criteria, a study design detail. |
| | | Population | Population | |
| | | Statistic | Variable | |
| | | Parameter | Study design detail | |

| ID | Verbatim response to RC task | Content element requested by RC Task | Code attributed to student response | Coding comments |
|---|---|---|---|---|
| 175 | what is the average amount of time a 10 year old boy watchs TV? (research question). A random sample of 500 boys aged 10 (sample) will be taken from schools across [country] (population). Then the results will be averaged to find the average amount of time (parameter) the boys watch TV. This number can than be applied to the population using statistical inference (statistic). | Sample | Sample | The response identifies an appropriate sample. Partial credit was awarded for population since it shows some understanding related to a pertinent sampling frame from which the sample could be drawn but is not sufficient to demonstrate a complete understanding of 'population.' The descriptions of 'parameter' and 'statistic' are approximately transposed. |
| | | Population | Partial | |
| | | Statistic | Parameter | |
| | | Parameter | Statistic | |
| 231 | In order to find out where people get their news from around the country, a survey could be conducted where 5,000 randomly selected people are told to name where they receive their news. They could be given choices such as TV, internet, newspaper, or magazines. Based on the results, since this is a sample of the entire country, it should be easy to tell where the entire country gets their news from. | Sample | Sample | The student identified an appropriate sample. Population is implied to be the whole country, but not stated explicitly, so partial credit was awarded. The student has not attempted to identify the statistic or parameter in the response. |
| | | Population | Partial | |
| | | Statistic | Not Identified | |
| | | Parameter | Not Identified | |

## 3.5 Analysis of response data

In order to address the research questions of the study, the data were scrutinized from several perspectives.

### *3.5.1 Identification of usable responses*

Each response was initially reviewed to determine whether it represented a genuine response to the RC task. Next, responses were evaluated to determine whether the student attempted to present a context or scenario of any kind. A third assessment identified whether a response included a conceivable research question or apparent line of inquiry.

### *3.5.2 Analysis of concepts attributed to the four target elements*

The first research question focuses on the aspects of students' invented context that are attributed to the targeted inferential modeling elements of sample, population, statistic and parameter. Analyses supporting the first research question included:

- Marginal distribution of response codes attributed to each target element;
- Cross-tabulation of scoring codes for the sample and population;
- Cross-tabulation of scoring codes for the statistic and parameter.

### *3.5.3 Analysis of conceptual diversity attributed to parameter*

The second research question focuses on determining if students incorrectly attribute a variety of concepts to the parameter and, if so, the nature of the incorrect concepts. Analyses supporting the second research question included:

- Cross-tabulation of response codes for the statistic and parameter among responses awarded at least partial credit for both the sample and population
- Cross-tabulation of response codes for the statistic and parameter among responses awarded full credit for both the sample and population

- Logistic regression of parameter response codes using response codes to sample, population and statistic as predictors where each content element was scored (1) if full or partial credit was earned or (0) if no credit was earned.

# 4. Results

## 4.1 Identification of usable responses

In total, 480 of 500 (96%) of responses were deemed genuine attempts. Responses deemed disingenuous (e.g., “use a calculator”) were excluded from further analysis. Responses that appeared genuine but were deemed irrelevant to the task (e.g., “I do not know how to begin to answer this question. I am so sorry for wasting your time”) were excluded. A total of 438 of 500 responses (87.6%) were found to provide a relevant attempt including a context or scenario as part of the solution.

## 4.2 Analysis of concepts attributed to the four target elements

The analysis considered the 438 responses that provided a context or scenario of some kind. Overall, 99 (22.6%) of the 438 attempts earned at least partial credit for identifying all four content elements (i.e. sample, population, statistic, and parameter). Furthermore, 71 responses (16.2%) earned full credit for clearly and correctly identifying the four content elements for their proposed context.

### *4.2.1 Marginal scoring distribution for concepts attributed to sample, population, statistic, and parameter*

Table 2 shows the distribution of responses that received full credit, partial credit, and no credit (all codes) for identification of the sample, population, statistic, and parameter among relevant attempts that provided a context or scenario of some kind as part of the solution. In general, students were more likely to receive full credit for the sample and population elements, and more likely to receive no credit for the statistic and parameter elements.

Table 2. Percent of students in each credit category within each content element for responses that provided a context (N = 438).

| Content Element | Credit Category<br>Full Credit | Partial Credit | Not Identified | Incorrectly Identified |
|---|---|---|---|---|
| Sample | 64.8 | 7.3 | 10.7 | 17.1 |
| Population | 58.9 | 8.2 | 18.7 | 14.2 |
| Statistic | 29.5 | 10.7 | 30.1 | 29.7 |
| Parameter | 20.3 | 8.2 | 36.5 | 34.9 |

Figure 2 shows the distribution and diversity of concepts inappropriately attributed to each of the target concept elements prompted by the RC task based on the 438 student responses that included a context. Responses awarded full credit or partial credit are not included in Figure 2 in order to more clearly illustrate the relative frequency of the remaining concepts. A total of 26 unique codes that represent content inappropriately attributed to the desired concept elements are referenced in Figure 2; code descriptions are available in Appendix 1.

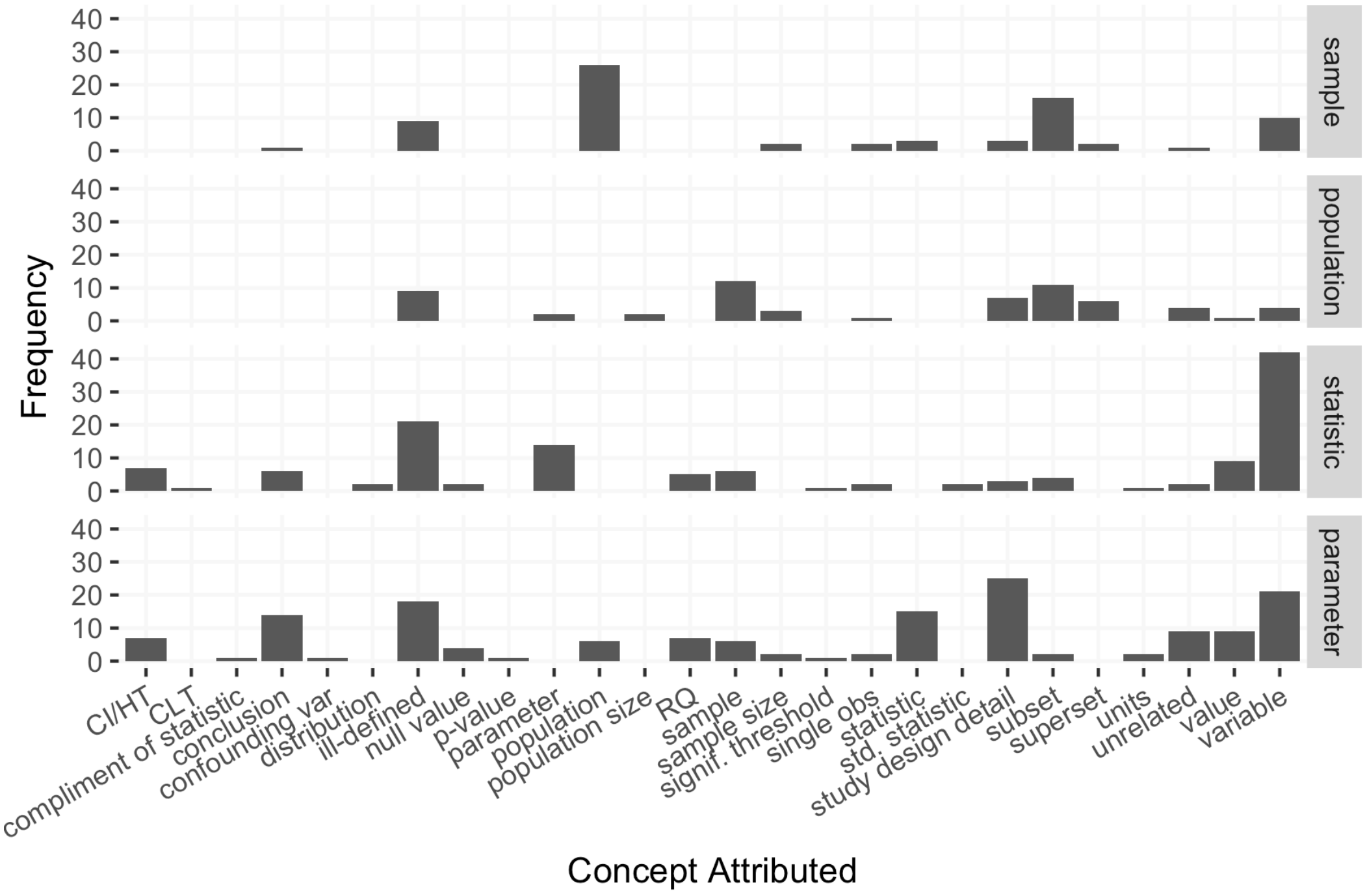

Figure 2. Frequency of incorrectly identified concepts attributed to each target element

As shown in Figure 2, 11 distinct codes summarize concepts incorrectly attributed to the sample. Common conceptions incorrectly attributed to the sample included descriptions of a population of some kind (not necessarily pertinent to the research question), an irrelevant subset, a variable among the data to be collected, or a response that was ill-defined/indiscernible. Additionally, 12 distinct codes summarize concepts incorrectly attributed to the population. Common conceptions incorrectly attributed to the population included descriptions of the sample, a proper subset of the target population, or a response that was ill-defined/indiscernible.

Incorrect conceptions commonly attributed to the statistic were distributed among 18 distinct codes. Most common among them included descriptions of a variable among the data to be collected, a response that was ill-defined/indiscernible, and descriptions of a parameter. Lastly, 20 distinct codes were used to summarize incorrect conceptions attributed to the parameter. The most frequent included a study design detail, a variable among the data to be collected, and a response that was ill-defined/indiscernible.

Figure 2 suggests the distributions for sample and population elements were similar, but distinct from the distributions for statistic and parameter, with the latter element pair having similar distributions. Analysis of the observed relationship was conducted using two new variables: one to represent the credit received on the sample/population element pair and a second to represent the credit received on the statistic/parameter element pair. Three credit categories were used for each element pair: Both Correct or Partial, Only One Correct or Partial, Both Incorrect (see Table 3).

Table 3. Credit category definitions for the content element pairs.

| Category | Description |
| --- | --- |
| Both Correct or Partial | Correct on both elements, correct on one element and partial on the other element, or partial on both elements |
| Only One Correct or Partial | Correct or partial on one element and either incorrect or no-attempt on the other element |
| Both Incorrect | Either incorrect or no-attempt on both elements |

Table 4 presents a cross-tabulation of the credit categories for the two element pairs. Receiving correct or partial credit on both concept elements of the sample/population element pair did not guarantee receiving credit on both concept elements of the statistic/parameter element pair, but it was more likely compared to receiving credit for only one of the elements or being incorrect on both the sample and population concept elements. The probability of being incorrect on both the statistic and parameter concept elements increased as the credit category for the sample/population element pair went from Both Correct or Partial to Only One Correct or Partial to Both Incorrect. A chi-square analysis produced a highly statistically significant result ($\chi^2(4) = 141.75$, $p < 0.001$; expected values for all cells > 18).

Table 4. Percent of students in each credit category for the statistic/parameter element pair within each credit category for the sample/population element pair (N = 438)

| | Statistic/Parameter | | |
| --- | --- | --- | --- |
| Sample/Population | Both Correct or Partial | Only One Correct or Partial | Both Incorrect |
| Both Correct or Partial | 39.3 | 19.0 | 41.7 |
| Only One Correct or Partial | 4.7 | 30.8 | 64.5 |
| Both Incorrect | 2.1 | 4.3 | 93.6 |

### 4.2.2 Concepts attributed to the sample and population

Figure 3 shows a complete cross-tabulation of coding for concepts attributed by each student to the sample and the population. Each cell in Figure 3 is an intersection of codes interpreting what the student directly attributed to “sample” on the x-axis and “parameter” on the y-axis. Each code corresponds to a vertical or horizontal grid line to trace each possible intersection. Magnitude is represented by a count within the cell and a grayscale shading analogous to a heat map with relatively darker shades representing larger magnitudes. Row and column totals appear in the margins and are similarly shaded to underscore frequency of each code used. Note that the above description applies similarly to Figures 4 and 5 below, and Figure 6 in Appendix 2.

Of the 438 responses, a majority (57.5%) earned at least partial credit for both the sample and the population. Roughly 14.6% earned at least partial credit for correctly identifying the sample but did not earn full or partial credit for the population. About 9.6% earned at least partial credit for correctly identifying the population but did not earn full or partial credit for the sample. Another 8.7% earned no credit for identifying the sample or population. Figure 3 shows additional detail specifying recognizable concepts that students attributed to the sample and population.

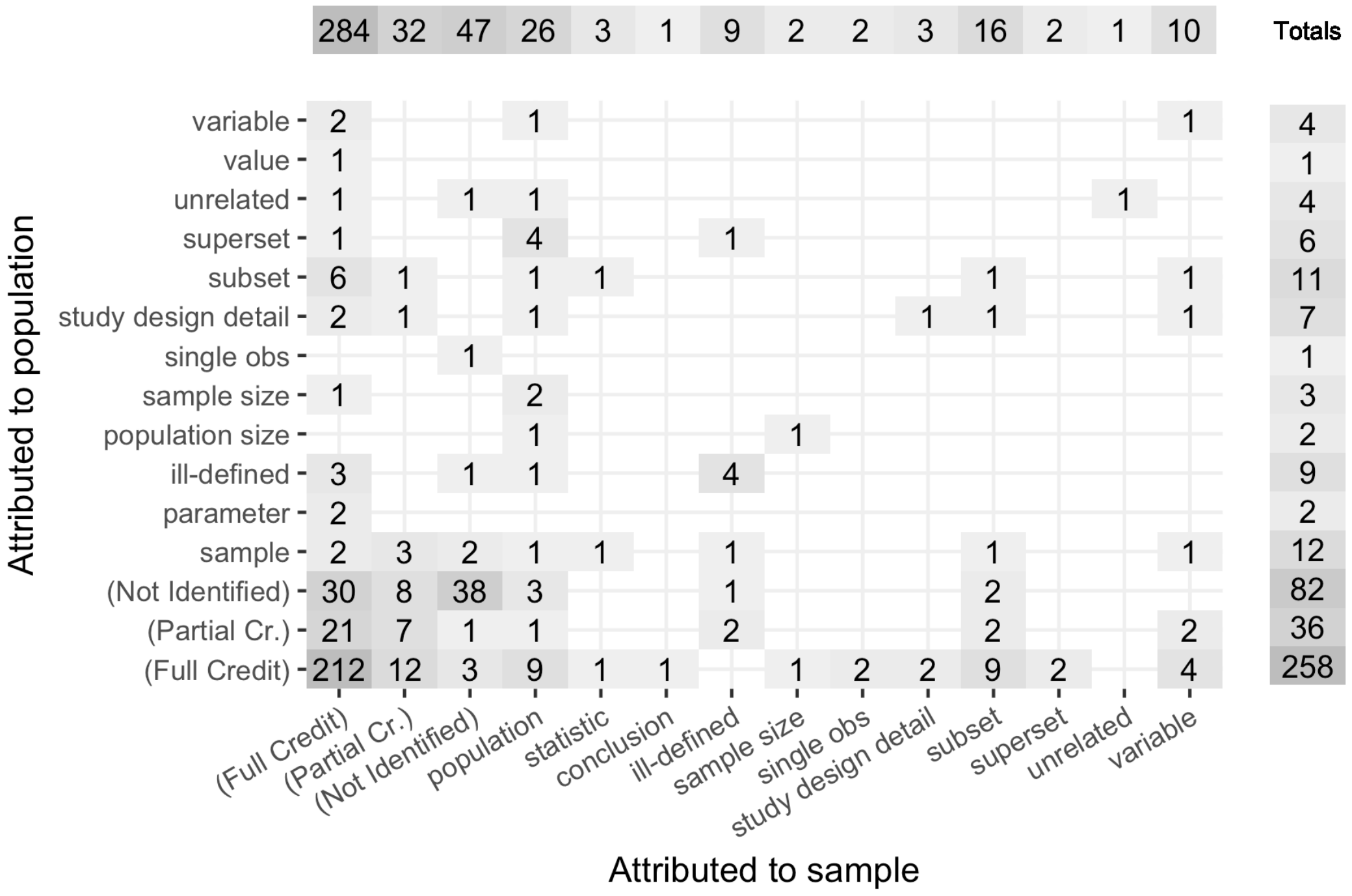

Figure 3. Cross-tabulation of concepts attributed to the sample and population (N = 438).

### 4.2.3 *Concepts attributed to the statistic and parameter*

Figure 4 shows a cross-tabulation of coding for concepts attributed by each student to the statistic and the parameter in their provided context. A total of 107 (24.4%) of the 438 responses earned at least partial credit for identifying both the statistic and the parameter. Another 69 students (15.8%) earned at least partial credit by correctly identifying the statistic, but not fully or partially identifying the parameter. By contrast, 18 students (4.1%) earned at least partial credit by correctly identifying the parameter, but not fully or partially identifying the statistic. Lastly, 113 students (25.8%) did not attempt to identify the statistic or parameter. Figure 4 shows additional detail including recognizable concepts that students attributed to the statistic and parameter.

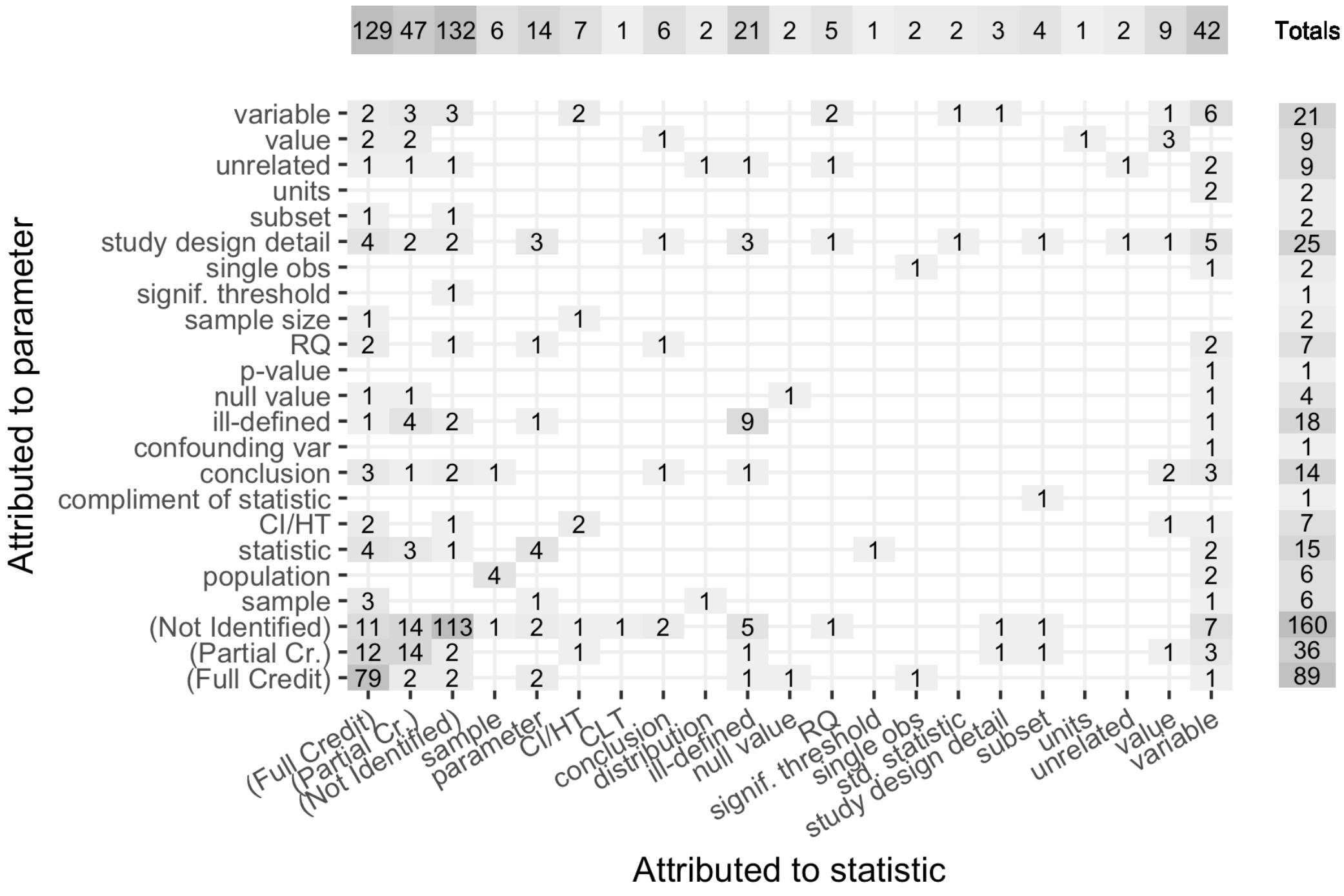

Figure 4 Cross-tabulation of concepts attributed to the statistic and parameter (N = 438).

## 4.3 Analysis of conceptual diversity attributed to parameter

The preceding analysis reported the diversity of responses attributed to the parameter (and statistic) without conditioning on identification of sample or population. The next two sections augment the analysis by investigation of response diversity attributed to the parameter and statistic after conditioning on identification of the sample and population, followed by a logistic regression analysis where identification of the parameter is conditioned on identification of the sample, population and statistic.

### *4.3.1 Cross-tabulation for the statistic and parameter conditioned on identification of the sample and population*

Figure 5 shows a cross-tabulation of codes for concepts attributed to the statistic and parameter by 252 students who earned partial or full credit for identifying both the sample and population. A total of 99 (39.3%) of the 252 students earned at least partial credit for identifying both the statistic and the parameter in their proposed context. Another 38 students (15.1%) earned at least partial credit for identifying the statistic but did not fully or partially identify the parameter. By contrast, 10 students (4.0%) earned at least partial credit for identifying the parameter but did not fully or partially identify the statistic. Lastly, 105 students (41.7%) earned no credit for the statistic or parameter elements.

Figure 5 shows additional detail specifying recognizable concepts that students attributed to the statistic and parameter. Common errors attributed to the parameter included a study design detail, a variable of some kind, an assumed numeric value, and the statistic. The most common error attributed to the statistic was description of a variable.

When the analysis is restricted to include only those students awarded full credit for both the sample and population (i.e. excluding 40 students awarded partial credit), the results are extremely similar. The analogous figure is included for the interested reader in Appendix 2.

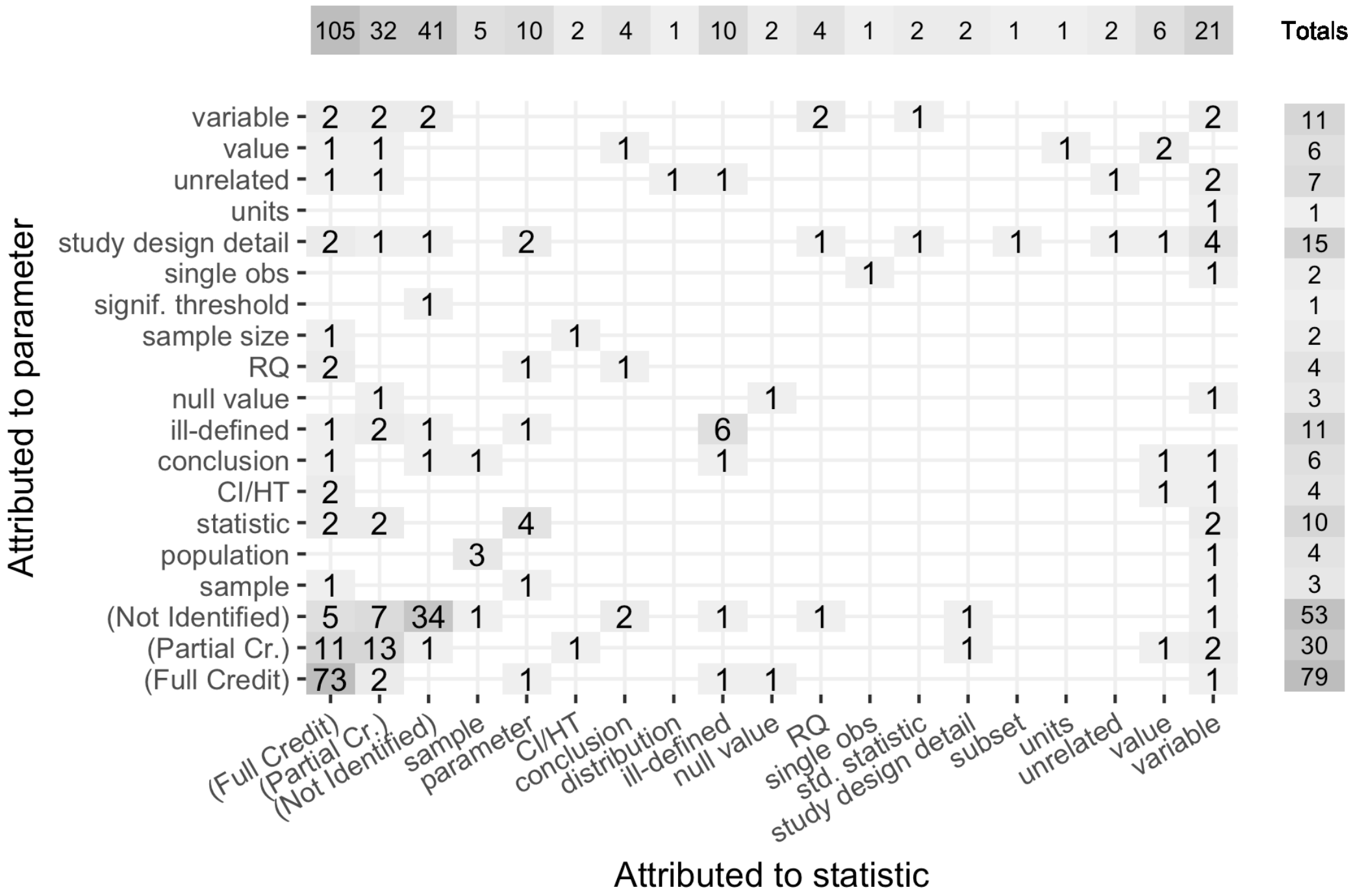

Figure 5. Cross-tabulation of concepts attributed to the statistic and parameter among responses with at least partial credit for both the sample and population (N = 252).

### *4.3.2 Logistic regression model of parameter given sample, population, and statistic.*

Because the data for scored descriptions of the sample, population, statistic, and parameter amount to four categorical variables with many possible outcomes, sparsity was problematic. To address the sparsity issue, responses associated with each content element were converted to binary outcomes based on whether or not at least partial credit had been earned. The resulting data are summarized in Table 5.

Table 5. Frequency of scoring patterns across sample, population, statistic, and parameter where "credit" indicates a response earning at least partial credit for an element (N = 438).

| | | | Parameter | |
|---|---|---|---|---|
| Sample | Population | Statistic | Credit | No Credit |
| Credit | Credit | Credit | 99 | 38 |
| Credit | Credit | No credit | 10 | 105 |
| Credit | No credit | Credit | 3 | 18 |
| Credit | No credit | No credit | 3 | 40 |
| No credit | Credit | Credit | 2 | 7 |
| No credit | Credit | No credit | 5 | 28 |
| No credit | No credit | Credit | 3 | 6 |
| No credit | No credit | No credit | 0 | 71 |

A logistic regression model was fit to the data using indicator variables for credit on the sample, population, and statistic elements to estimate the odds of success when identifying the parameter in a response to the RC task.

$$\log\left(\frac{p}{1-p}\right) = \beta_0 + \beta_1(Sample) + \beta_2(Population) + \beta_3(Statistic)$$

Larger models including two-way and three-way interaction terms were also evaluated. The full model with all main effects and interactions over-fitted the data with terms for all 8 possible outcomes associated with the explanatory variables. Consequently, a more appropriate comparison evaluated the improvement of augmenting the additive model with two-way interactions (i.e., an interaction model). A drop-in-deviance test comparing the additive model and the interaction model was not statistically significant ($p$ = 0.644). Both AIC and BIC indices favored the additive model and residual deviance associated with the additive model did not suggest significant lack-of-fit. The model summary and coefficient estimates for the additive model are shown in Table 6.

Table 6. Summary of additive model fit for logistic regression of successful identification of the parameter conditional on identification of sample, population, and statistic.

| | Estimate | exp(Estimate) | Std. Error | z value | $p$-value |
|---|---|---|---|---|---|
| (Intercept) | -4.3541 | 0.013 | 0.5091 | -8.552 | < 0.001 |
| Sample | 0.5384 | 1.713 | 0.4296 | 1.253 | 0.210 |
| Population | 1.8191 | 6.166 | 0.4117 | 4.419 | < 0.001 |
| Statistic | 2.8035 | 16.502 | 0.3023 | 9.275 | < 0.001 |

According to Table 6, the odds that a response to the RC task earned full or partial credit for properly identifying the parameter decreased by nearly 99% if the response did not also earn credit for at least one other content element among the sample, population, and statistic. By contrast, the odds that a response to the RC task earned credit for properly identifying the parameter increased more than 16-fold if it also earned credit for identifying the statistic. Similarly, the odds that a response to the RC task earned credit for properly identifying the parameter increased more than 6-fold if it also earned credit for identifying the population. Although the sample indicator variable is not statistically significant in the additive model, this may be due to multicollinearity as it becomes statistically significant if the indicator variable for either population or statistic is removed from the model.

## 5. Discussion

Relating a statistic and parameter to a context of their choosing was more difficult for students than identification of the sample and population. This result speaks directly to the challenge of undergraduate introductory statistics students as they develop schema for statistical modeling. Defining "statistical model" to mean a statistical approximation designed to reflect or describe the underlying structure of a data generating process, responses to the RC task suggest several implications.

## 5.1 Utility of the RC task

At its core, the RC task could be likened to a definitions task. Rather than requiring students to produce a rote definition of each concept in general terms, the RC task requires them to place their definition in context which is not unlike asking grade school students to "use it in a sentence" when learning new vocabulary words. In the latter analogy, a student may consider a few possible sentences and select one that best displays their understanding of the vocabulary term. This act of self-critique in which students align their understanding of the elements of inferential modeling is a hallmark of statistical thinking (Chance 2002; Wild and Pfannkuch 1999) and is essential for successful cognitive transfer (e.g., Bransford et al. 2000; Singley and Anderson 1989). In response to the RC task, students could certainly choose a different context if they truly understood the content elements but struggled at first to adequately demonstrate understanding within the context that first came to mind. In fact, this practice was observed by the first author during at least one cognitive interview with a student piloting the RC task. A student's choice implies that the provided context was the best or most complete illustration of sample, population, statistic, and parameter that the student was able to put forth in that moment.

## 5.2 Concepts attributed to sample, population, statistic, and parameter

According to the results, students were more frequently successful when describing the sample and population when compared to the statistic and parameter. By contrast, students were more likely to omit an attempt to identify the statistic and/or parameter from their response and showed far more diversity of concepts incorrectly attributed to the statistic and parameter. This diversity of concepts may be due to a lower level of emergent understanding that results in describing some other familiar concept from their statistics course.

Despite such wide diversity of concepts attributed to each target element, a few patterns did emerge, particularly among responses associated with the sample, statistic, and parameter. Although students were often able to appropriately identify the sample in their proposed context, it is perhaps not surprising that the population was the most common concept incorrectly attributed to the sample. By contrast, the statistic and parameter were apparently more challenging for students to accurately describe. Among students that did not earn full or partial credit for their description of the statistic, responses commonly described a variable or data to be collected, observed, or measured without acknowledging the role of the statistic as a point estimate, summary measure, aggregation, etc. of sample data (e.g., Table 1, ID 177). This pattern of attributing a variable of some kind to the statistic was the most pronounced and disproportionately common misconception, occurring in nearly 10% (42 of 438) of all scored responses that provided a context (see Figure 4).

### 5.3 Conceptual diversity attributed to parameter

An interesting pattern also emerged among descriptions attributed to the parameter such that the most common response pattern among those that did not earn at least partial credit included description of study design detail (e.g., Table 1, ID 148: "parameter: one semester"). Apparently, this revealed a source of confusion between the definition of a statistical parameter as it relates to a structural feature of some data-generating process, and a conflicting use of the term to describe a constraint or condition imposed during data collection. This independently corroborates concurrent work by Kaplan and Rogness (2018) that described a similar pattern among student responses equating the term parameter to "a rule, characteristic, or condition for inclusion or a limit or boundary" (p. 8).

The diversity of contexts and definitions provided by the students for the parameter element suggests that lexical ambiguity about the statistical meaning of the word "parameter" may need to be addressed. Given the diversity of meanings presented by students in the current study, developing a correct statistical understanding of the word "parameter" may be challenging. In general, one approach to helping students develop a correct understanding for the statistical usage of a word is to leverage students' initial or colloquial meanings to create opportunities for students to compare and contrast the every-day and technical meanings (Kaplan et al. 2010).

## 5.4 Relationship between statistical inference and model recognition

Statistical inference is simply one of many ways that a statistical model can be utilized. In addition to inference, a statistical model may be used for purposes such as exploratory data analysis (EDA), estimation, and prediction. According to a non-scientific sampling of introductory statistics textbooks immediately accessible to the authors, it seems that most introductory statistics courses and textbooks primarily discuss different uses of statistical modeling and very rarely discuss the unifying construct of statistical modeling itself. Since productive EDA, inference, estimation, and prediction are predicated on the statistical modeling construct, students with a low level or disconnected understanding may retreat to rote memorization of technical jargon or software steps when pressed to display deeper understanding. This tendency was not uncommon among responses to the RC task. For example, when surveying the cross-tabulations associated with concepts attributed to the statistic and the parameter, there were many instances in which the student described a *p*-value, null hypothesis, or steps of constructing a confidence interval or performing a hypothesis test with a calculator or software. Many of these could be coded for use in the cross-tabulation, but more extreme cases sometimes included an amalgam of unrelated statistical jargon that was coded as "ill-defined." The latter was coded if there was clear attribution to either the sample, population, statistic, or parameter, yet others were coded "not identified" when no apparent attribution to a target content element was found.

While previous and concurrent research suggested identification of the parameter in an invented context to be difficult for introductory statistics students (e.g., Beckman 2015; Kaplan and Rogness 2018), it was somewhat surprising that the statistic was similarly challenging. In the frequentist paradigm of inferential statistics, the parameter is an abstract concept representing an unknown feature of a population. By comparison the statistic is more tangible as a calculated summary of observed data. For those reasons, one might expect the statistic to be the centerpiece of many discussions in an introductory statistics course and more relatable to students. This too could be an indicator of the challenge facing students as they develop understanding of statistical modeling. When defining a "statistical model" there is some room for debate about whether the term refers to the theoretical or the empirical (e.g. Garfield and Ben-Zvi 2008; Graham 2006; McCullagh 2002). Perhaps this distinction is unimportant since students' difficulty with interpreting both the statistic and parameter suggests that neither the empirical view nor the theoretical model is more widely relatable to students.

## 5.5 Limitations

One limitation of the data analysis relates to scoring descriptions of sample, population, statistic, and parameter as though they are concepts that could be understood independently, when they are clearly not. Furthermore, descriptions attributed to these concepts cannot be assumed independent. For example, it was quite unusual for a student to attribute the same description to two—much less three or four—of the requested content elements (e.g., an identical description attributed to both the sample and the population) as might be expected occasionally if descriptions of each content element were truly independent. It isn't possible to know what students were thinking or how frequently they were torn between two or more labels when describing the four requested content elements, so there was no attempt to accommodate for this behavior during data analysis.

No covariates related to student demographics or achievement data were collected and there was never an attempt of any kind to standardize, control, or otherwise influence curriculum delivery. Importantly, this could mean that the term "parameter" may not be used regularly (if at all) by some courses or instructors. Furthermore, generalizability of the sample may be questionable since participants were recruited by soliciting instructor volunteers from several large email lists of statistics educators predominantly, though not exclusively, affiliated with institutions in the United States and Canada. Among those reached, however, the response rate evaluated as a proportion of relevant responses or as a proportion of responses with a viable context included were 96% and 87.6%, respectively, so non-response does not appear to be a serious concern.

Lastly, there was no method or attempt to evaluate whether students actually provided a unique context of their own invention rather than something from a textbook or Internet search. Responses very rarely appeared to utilize suspiciously similar contexts and structure.

## 5.6 Implications for Teaching, Assessment, and Research

Because statistical inference is predicated on a statistical model, perhaps students would benefit from teaching and assessment practices that regularly highlight this relationship and frame inferential methods within a larger modeling context.
As for teaching practices, the concept of "parameter" plays a critical role in developing deep understanding of statistical modeling and its various uses. Introductory statistics textbooks, and likely courses, seem to emphasize a line of thinking that a sample comes from some population and then one or more statistics are calculated from the collected data. This is problematic because omitting the role of the parameter leaves the logic of inference incomplete. As a result, we would do well to augment this line of thinking that a sample comes from some population and then we calculate one or more statistics from the collected data *in order to estimate a parameter*. The parameter is then a central purpose rather than an abstract afterthought. This idea similarly extends to the statistical model where inference is recast explicitly as a *use* of the statistical model, so the statistical model is not viewed as a separate construct reserved for topics and courses beyond the scope of an introductory statistics course.

In assessment, innovative tasks like the RC task that require students to provide a context of their own invention should be considered for inclusion in the assessment strategy for statistics courses. Such tasks reveal a different perspective on student understanding and allow the instructor to quite easily identify students who seem to know the statistical jargon without deeper understanding. One possible argument against use of an invented context task like the RC task might be that there is a risk that students describe a scenario or context in which some of the desired content elements are not present.

For example, several students described applied probability scenarios for which they would estimate the probability of observing an event from an assumed probability distribution (usually Normal). There was no relevant sample or statistic for the student to identify, so one might argue that the students were disadvantaged by their chosen scenarios. However, the choice of an applied probability context may reveal an incomplete understanding of important content elements and the assessment task can be claimed to have accomplished its purpose. A context provided by the instructor can include all of the desired content elements that the student is to identify, but this may invite bias if students benefit from a process of elimination or other content present in the task which could result in an overestimate of student understanding. Variations on the RC task could also be used as part of instruction, such as asking students to identify the missing elements in a context description or presenting applied probability contexts and asking students to discuss which of the desired elements are applicable, which are not, and why.

This study attempts to address an apparent gap in the statistics education research literature base by contributing preliminary evidence of students' model recognition ability. Additional research is recommended to study student outcomes when a statistical modeling framework is emphasized in the introductory statistics curriculum. Furthermore, a thorough analysis and categorization of all the different ways in which students "defined" parameter and statistic would be useful to better characterize and address associated misconceptions. Lastly, the research questions posed by students may represent an interesting line of inquiry.

While primary school students do not naturally pose statistical questions, there is evidence that their ability to do so can be developed through guided inquiry activities (Allmond and Makar 2010). Not much is known about college students' ability to pose statistical questions. Students' ability to pose suitable statistical questions may well affect their ability to identify the statistic and parameter in a context.

**Acknowledgements**

The authors wish to express their sincere gratitude to Joan Garfield for her thoughtful direction and influence on development of early research leading to this work. The authors are also grateful for the thoughtful feedback and constructive suggestions of colleagues during the SRTL-10 forum.

# References

Allmond, S., & Makar, K. (2010). Developing primary students' ability to pose questions in statistical investigations. In C. Reading (Ed.), *Data and context in statistics education: Towards an evidence-based society. Proceedings of the 8th international conference on teaching statistics*. Voorburg, The Netherlands: International Statistical Institute.

Beckman, M. D. (2015). *Assessment of cognitive transfer outcomes for students of introductory statistics* (Doctoral dissertation, University of Minnesota—Twin Cities). Retrieved from http://iase-web.org/documents/dissertations/15.MatthewBeckman.Dissertation.pdf

Ben-Zvi, D., & Garfield, J. (2004). Statistical literacy, reasoning, and thinking: Goals, definitions, and challenges. In D. Ben-Zvi & J. Garfield (Eds.), *The challenge of developing statistical literacy, reasoning and thinking* (pp. 3-15). Dordrecht, The Netherlands: Kluwer Academic Publishers.

Bransford, J. D., Brown, A. L., & Cocking, R. R. (2000). *How people learn: Brain, mind, experience, and school: Expanded edition.* Washington, DC: National Academies.

Chance, B. (2002). Components of statistical thinking and implications for instruction and assessment. *Journal of Statistics Education, 10*(3). Retrieved from http://ww2.amstat.org/publications/jse/v10n3/chance.html

delMas, R., Garfield, J., Ooms, A., & Chance, B. (2007). Assessing students' conceptual understanding after a first course in statistics. *Statistics Education Research Journal, 6*(2), 28-58.

Garfield, J., & Ben-Zvi, D. (2008). *Developing students' statistical reasoning: Connecting research and teaching practice*. Springer Science & Business Media.

Graham, A. (2006). *Developing thinking in statistics*. London: Paul Chapman.

Haller, H., & Krauss, S. (2002). Misinterpretations of significance: A problem students share with their teachers? *Methods of Psychological Research, 7*(1), 1-20.

Kaplan, J. J., Fisher, D. G. & Rogness, N. T. (2009). Lexical ambiguity in statistics: What do students know about the words association, average, confidence, random and spread? *Journal* *of Statistics Education, 17*(3). Retrieved from www.amstat.org/publications/jse/v17n3/kaplan.html

Kaplan, J. J., Fisher, D. G. & Rogness, N. T. (2010). Lexical ambiguity in statistics: How students use and define the words: association, average, confidence, random and spread. *Journal of Statistics Education, 18*(2). Retrieved from www.amstat.org/publications/jse/v18n2/kaplan.pdf

Kaplan, J.J. & Rogness, N. (2018). Increasing statistical literacy by exploiting lexical ambiguity of technical terms. *Numeracy, 18*(1), 1-14.

Lavigne, N. C. & Lajoie, S. P. (2007). Statistical reasoning of middle school children engaging in survey inquiry. *Contemporary Educational Psychology, 23*(4), 630-666.

Makar, K., & Rubin, A. (2009). A framework for thinking about informal statistical inference. *Statistics Education Research Journal, 8*(1), 82-105.

McCullagh, P. (2002). What Is a Statistical Model? *The Annals of Statistics, 30*(5), 1225-1267. Retrieved from http://www.jstor.org/stable/1558705

Meletiou-Mavrotheris, M. & Paparistodemou, E. (2015). Developing students' reasoning about samples and sampling in the context of informal inferences. *Educational Studies in Mathematics, 88*(3), 385-404.

Pfannkuch, M. (2006). Informal inferential reasoning. In A. Rossman & B. Chance (Eds.), *Working cooperatively in statistics education. Proceedings of the 7th international conference on teaching statistics.* Voorburg, The Netherlands: International Statistical Institute.

Pfannkuch, M., Ben-Zvi, D., & Budgett, S. (2018). Innovations in statistical modeling to connect data, chance, and context. *ZDM Mathematics Education.* (**this issue**)

R Core Team (2017). *R: A language and environment for statistical computing*. R Foundation for Statistical Computing, Vienna, Austria. URL https://www.R-project.org/

Reed, S. K., Dempster, A., & Ettinger, M. (1985). Usefulness of analogous solutions for solving algebra word problems. *Journal of Experimental Psychology: Learning, Memory, and Cognition, 11*(1), 106-125.

Rossman, A. J., & Chance, B. L. (2001). *Workshop statistics: Discovery with data (2nd ed.).* Emeryville, CA: Key College Publishing.

Singley, M. K., & Anderson, J. R. (1989). *The transfer of cognitive skill.* Cambridge, MA: Harvard University Press.

Vallecillos, A. (1999). Some empirical evidence on learning difficulties about testing hypotheses. *Bulletin of the International Statistical Institute: Bulletin of the 52nd Session of the International Statistical Institute, 58*, 201-204.

Watson, J. M., & Kelly, B. A. (2005). Cognition and instruction: Reasoning about bias in sampling. *Mathematics Education Research Journal, 17*(1), 25-27.

Watson, J. M., & Moritz, J. B. (2000). Development of understanding of sampling for statistical literacy. *The Journal of Mathematical Behavior, 19*(1), 109-136.

Well, A.D., Pollatsek, A., & Boyce, S.J. (1990). Understanding the effects of sample size on the variability of the mean. *Organizational Behavior and Human Decision Processes, 47*, 289–312.

Wild, C. J., & Pfannkuch, M. (1999). Statistical thinking in empirical enquiry. *International Statistical Review, 67*(3), 223-248.

Williams, A.M. (1999). Novice students' conceptual knowledge of statistical hypothesis testing. In J.M. Truran, & K.M. Truran (Eds.), *Making the difference: Proceedings of the twenty-second annual conference of the mathematics education research group of Australasia* (pp. 554-560). Adelaide, South Australia: MERGA.

# Appendix 1: Rossman-Chance (RC) Task Scoring Guidance

The complete guidance document with remarks to accompany each prompt encountered while scoring the RC Task is available in the online supplement. A list of codes used and an accompanying description is reproduced here.

### Scoring codes applied to the RC task

| Code | Description |
|---|---|
| “sample” (or “1”) | the requested element (i.e. sample, population, statistic, parameter) was identified as the “sample” in the response. |
| “population” (or “2”) | the requested element (i.e. sample, population, statistic, parameter) was identified as the “population” in the response. |
| “statistic” (or “3”) | the requested element (i.e. sample, population, statistic, parameter) was identified as the “statistic” in the response. |
| “parameter” (or “4”) | the requested element (i.e. sample, population, statistic, parameter) was identified as the “parameter” in the response. |
| “none” (or “0”) | student did not attempt to identify the requested element (i.e. sample, population, statistic, parameter) in the response or simply stated a definition |
| “full credit” | not used during coding; this designation was applied during analysis to clarify that the requested element was clearly and correctly identified in the response. |
| "partial" | indicates that provided text is insufficient for full |

| Code | Description |
|---|---|
| | credit, but response shows evidence of (likely) understanding |
| "variable" | indicates that student appears to describe one or more variables in the study |
| "value" | is an assumed number (e.g. for statistic or parameter) without sufficient explanation |
| "conclusion" | indicates that student appears to draw a conclusion or state some result for the research question |
| "ill-defined" | indicates that student explicitly attributes something to the sample/population/statistic/parameter but it isn't clear how it relates or what relevant role it plays |
| "subset" | indicates that student overtly describes some subset of the population/sample/etc that could not itself be used to address the research question (i.e. sample/population for an RQ about the proportion that say "yes" to a survey question can't be restricted to a subset who say "yes" only) |
| "superset" | indicates that student overtly describes some superset of the population/sample/etc that does not directly address the research question (i.e. population of interest is "STAT 100 students" superset might be "university students") |
| "unrelated" | e.g. indicates that the student overtly specifies a sample/population that is not related to the research question, or other unrelated detail |
| "single obs" | student attributes one specific observation from the data to the sample/statistic/etc |

| Code | Description |
|---|---|
| "RQ" | student restates all or part of the research question |
| "study design detail" | describes some specific detail (e.g. duration, condition, selection criterion) of the study |
| "sample size" | student attributes the sample size to the target concept |
| "CI/HT" | e.g., student claims the parameter is a confidence interval |
| "signif. threshold" | student explicitly attributes a significance threshold such as $\alpha = 0.05$ |
| "p-value" | student explicitly attributes a p-value |
| "null value" | some value hypothesized for a parameter of interest to be tested |
| "units" | unit of measurement (e.g. pounds) |
| "std. statistic" | e.g., z-score, test statistic |
| "confounding var" | student describes a confounding variable that could impact the response described |
| "distribution" | student attributes a specific distribution (e.g. Normal distribution; t-distribution) |

# Appendix 2: Concepts attributed to the statistic and parameter among responses with full credit for both the sample and population

Figure 6 presents a cross-tabulation of codes for concepts attributed to the statistic and the parameter by 212 students who earned full credit for both the sample and population elements.

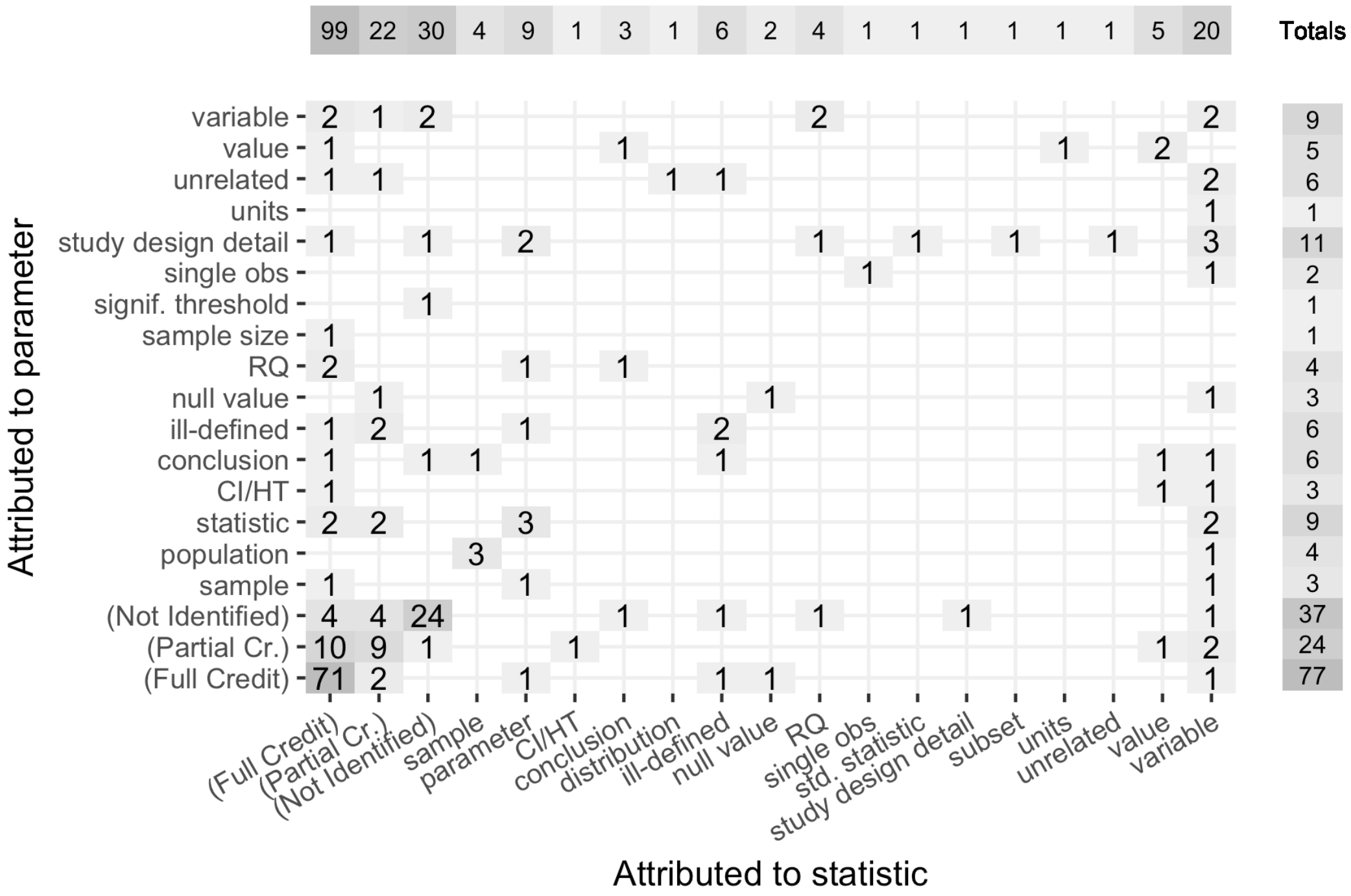

Figure 6. Cross-tabulation of concepts attributed to the statistic and parameter among responses with full credit for both the sample and population (N = 212).

A total of 92 (43.4%) of the 212 students earned at least partial credit for identifying both the statistic and the parameter. Another 29 students (13.7%) earned at least partial credit for identifying the statistic but did not earn credit for the parameter. By contrast, 9 students (4.2%) earned at least partial credit for identifying the parameter but did not earn credit for the statistic. Lastly, 82 students (38.7%) earned no credit for the statistic or parameter, despite earning full credit for both the sample and population. Figure 6 shows additional detail specifying recognizable concepts attributed to the statistic and parameter. Common errors attributed to the parameter included study design detail, a description of a variable, or the statistic. Again, the most common error attributed to the statistic was description of a variable.